\documentclass[10pt, journal, letterpaper]{IEEEtran}

\usepackage{cite}
\usepackage{amsmath,amssymb,amsfonts}
\usepackage{algorithm}
\usepackage{algorithmic}
\usepackage{graphicx}
\usepackage{textcomp}
\usepackage{subfigure}
\usepackage{setspace}
\usepackage{longtable}
\usepackage{multirow}
\usepackage{enumerate}
\usepackage{color}
\usepackage{dsfont}

\allowdisplaybreaks[2]

\def\BibTeR{{\rm B\kern-.05em{\sc i\kern-.025em b}\kern-.08em
    T\kern-.1667em\lower.7ex\hbox{E}\kern-.125emR}}

\begin{document}

\title{
MIPS: Instance Placement for Stream Processing Systems based on
Monte Carlo Tree Search
}

\author{Xi Huang, Ziyu Shao$^*$, Yang Yang
\IEEEauthorblockA{\\School of Information Science and Technology, ShanghaiTech University\\
 Email: \{huangxi, shaozy, yangyang\}@shanghaitech.edu.cn}
 \thanks{
 	$^*$ The corresponding author of this work is Ziyu Shao.
 }}

\maketitle

\begin{abstract} 
	Stream processing engines enable modern systems to conduct large-scale analytics over unbounded data streams in real time.  They often view an application as a direct acyclic graph with streams flowing through pipelined instances of various processing units. One key challenge that emerges is \textit{instance placement}, \textit{i.e.}, to decide the placement of instances across servers with minimum traffic across servers and maximum resource utilization. 
	The challenge roots in not only its intrinsic complexity but also the impact between successive application deployments. Most updated engines such as Apache Heron exploits a more modularized scheduler design that decomposes the task into two stages: One decides the instance-to-container mapping while the other focuses on the container-to-server mapping that is delegated to standalone resource managers.
	The unaligned objectives and scheduler designs in the two stages may lead to long response time or low utilization. However, so far little work has appeared to address the challenge.
	Inspired by recent success of applications of Monte Carlo Tree Search (MCTS) methods in various fields, we develop a novel model to characterize such systems, formulate the problem, and cast each stage of mapping into a sequential decision process. By adopting MCTS methods, we propose \textit{MIPS}, an {M}CTS-based {I}nstance {P}lacement {S}cheme to decide the two-staged mapping in a timely yet efficient manner. 
	In addition, we discuss practical issues and refine MIPS to further improve its performance. Results from extensive simulations show, given mild-value of samples, MIPS outperforms existing schemes with a significant traffic reduction and utilization improvement. 
	To our best knowledge, this paper is the first to study the two-staged mapping problem and to apply MCTS to solving the challenge.
\end{abstract}

\section{Introduction}

Recent years have witnessed an explosive growth of data streams that are incessantly generated from a wide assortment of applications, \textit{e.g.}, Twitter\cite{storm}, Facebook\cite{facebook}, and LinkedIn\cite{le2012linked}.
To conduct large-scale, real-time processing for such data streams, 
a number of stream processing engines have been proposed and launched, aiming at high scalability, availability, responsiveness, and fault tolerance\cite{samza, flink, le2012linked, storm, herondoc, facebook}.

To date, stream processing engines have evolved over three generations\cite{heinze2014cloud}. 
Among most recent third-genernation engines, Apache Storm\cite{storm} and Heron\cite{herondoc} stand out by their extensive adoption and support from a large community\cite{stormcomm}\cite{heroncomm}, as well as their modularized and scalable design. 
%Originally initiated by Nathan Marz at BackType in 2010, Apache Storm was acquired by Twitter in 2011 and later open sourced in 2012.
Typically, engines like Storm and Heron  view each stream processing application as a direct-acyclic graph, \textit{a.k.a.} a topology, where data streams (edges) are processed through pipelined components (nodes).   
%A topology has two types of components, \textit{spouts} and \textit{bolts}. Spouts are source nodes that read data streams in from external data sources or messaging queues\cite{kafka}, pre-process and generate tuples fed to subsequent nodes. 
%Meanwhile, bolts receive and process tuples from their predecessors, then persist or forward consequent tuples to their successors.

\begin{figure}[h!]
	\centering
	\subfigure[A potential trade-off between traffic reduction and high resource utilization]{ 
    	\includegraphics[scale=.33]{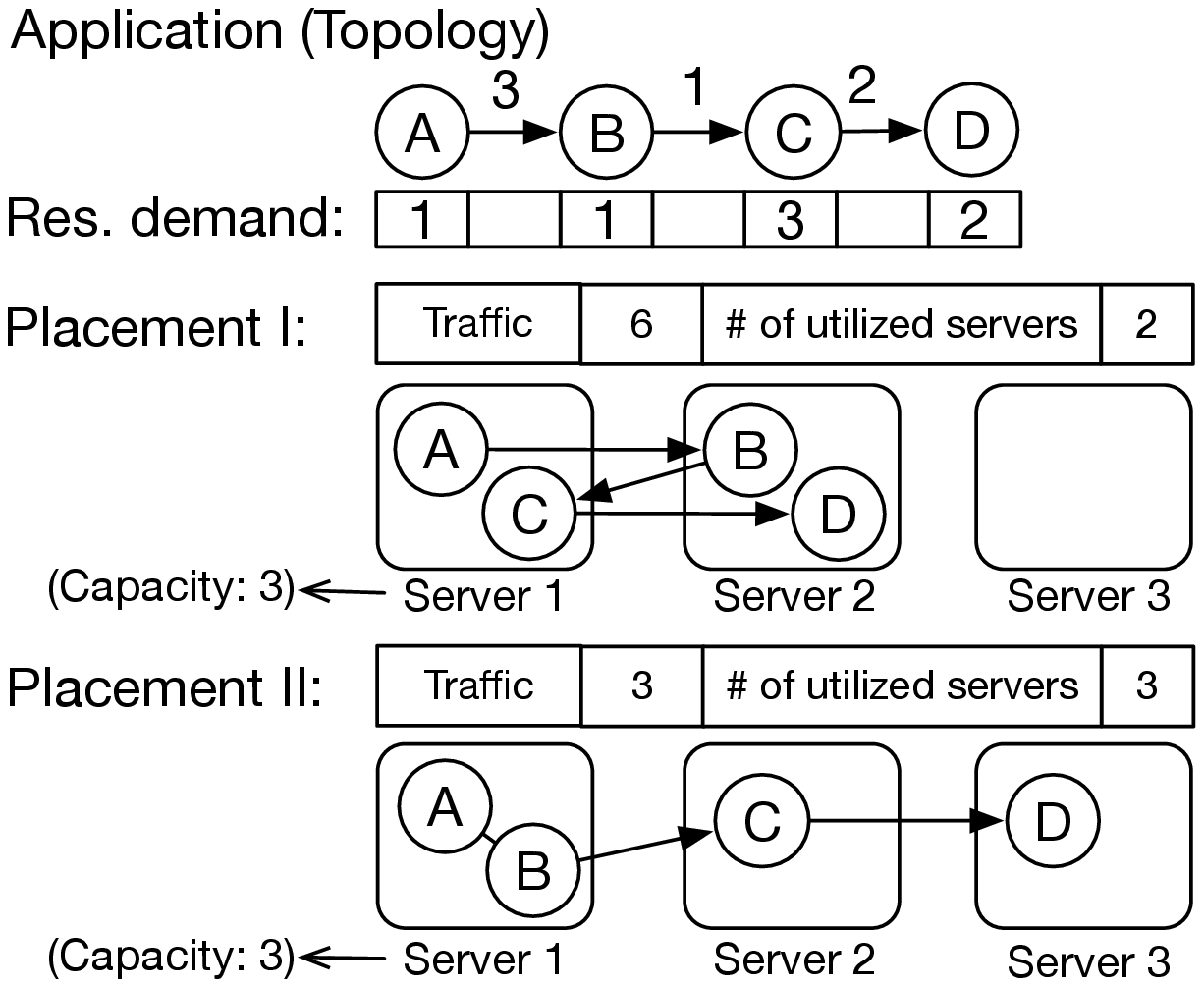}
  	}
	\subfigure[Instance placement in two stages with discrepant goals]{ 
    	\includegraphics[scale=.33]{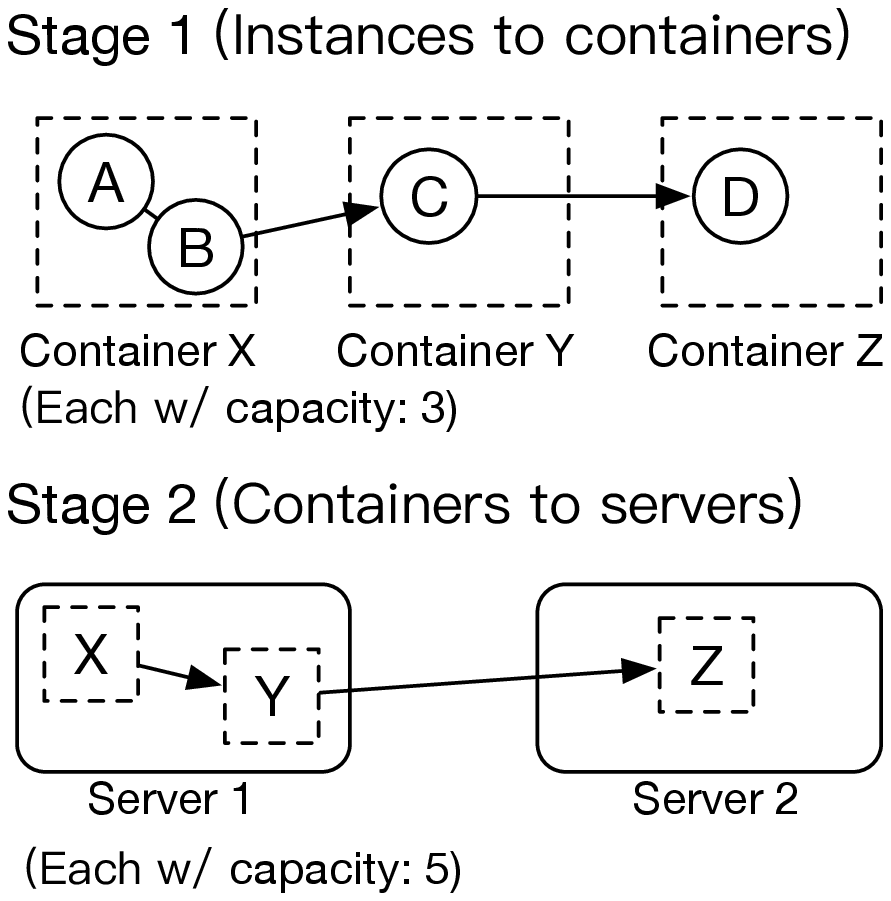}
  	}
	\caption{
	\textit{Basic settings:} A topology consists of four components, each with one instance. 
	\textit{Observations:}
	(a) Placement I incurs $6$ units of traffic between the two utilized servers, while Placement II reduces the traffic by $3$ units but at the cost of one extra server being used.
	(b) A sample two-staged instance placement: Instances are distributed with minimum cross-container traffic, then containers are deployed to servers in the order of (X, Y, Z), each assigned to the server with minimum but sufficient resources (best fit).   
	}
	\label{example1}
\end{figure}

To launch applications, users submit their requests to the system scheduler.   
Requests arrive in an online manner, each
specifying the topology of a given application, along with the \textit{parallelism} requirement, \textit{i.e.}, the number of instances for each component, and their resource demands.
Upon deployment, a key step and challenge to the scheduler is \textit{instance placement}, 
\textit{i.e.}, to decide how to distribute instances 
within a cluster of heterogeneous servers, containers\cite{docker}, or processes. 
Besides the intrinsic complexity of the problem being $\mathcal{NP}$-hard\cite{peng2015r}, 
instance placement is often associated with two objectives: 
1) to shorten response time, the instances of successive components should be placed in proximity with minimum cross-server or -container traffic;
2) to achieve high resource utilization, the placement should utilize as few servers/containers as possible. 
However, these two objectives may conflict each other, as Figure \ref{example1}(a) shows. 
Any inadvertent placement could lead to either high cross-server traffic with long response time, 
or low resource utilization with unnecessary overheads.

Constrained by the underlying scheduler design, instance placement generally falls into two categories. 
The first one is typified by Storm. 
Terming each instance a task, 
Storm's scheduler manages both the computation and enforcement of task placement with a direct control over its underlying cluster's resources \cite{storm}. 
Consequently, the scheduler can directly map tasks to servers through an one-staged decision, each server running a few processes that host tasks within their threads.
However, Storm's built-in schemes, \textit{round-robin} (RR) and \textit{first-fit-decreasing} (FFD), are often blamed for their blindness to traffic patterns between components upon decision making \cite{p-scheduler}. 
Motivated by that, previous studies abound, focusing on designing one-staged placement schemes\cite{peng2015r,t-storm,aniello2013adaptive,p-scheduler}.
Despite their effectiveness, such an integrated scheduler design has come to its end due to the highly coupled implementation of resource management and scheduling\cite{kulkarni2015twitter}.

\setlength{\textfloatsep}{0pt} 
This calls for a more modularized scheduler design, which typifies the second category.
One landmark is Twitter's replacing Storm with Heron in 2015\cite{herondoc}.
Compared to Storm, Heron refines the engine design by treating each instance as an independent process in a container, 
while delegating resource management to more established cluster schedulers such as YARN\cite{yarn} and Nomad\cite{nomad}.
Consequently, instance placement is decided in two stages. 
The first stage focuses on distributing instances onto a set of containers, \textit{a.k.a.}, \textit{instance-container mapping}, often done by  Heron's scheduler.
Next, containers are submitted as jobs and assigned to servers by the cluster scheduler, \textit{a.k.a.},
\textit{container-server mapping}.
Compared to one-staged schemes, the design for two-staged placement is even more challenging: 
Any unilateral effort in optimizing the placement on either stage would be in vain if their objectives are inconsistent.
Figure \ref{example1} (b) shows such an example: Stage $1$ decides the instance-container mapping with minimum traffic across containers,
whereas Stage $2$ focuses on maximizing resource utilization by using \textit{best fit} heuristic, inducing $2$ units of traffic between containers. However, the optimum amounts of traffic is $1$ by placing container Y, Z together in the same server and container X in the other. This is due to stage 1 aims at reducing cross-container traffic whereas stage 2 doesn't. 
By far, very little work has been conducted to work on this two-staged mapping problem. 
Often times solutions for Storm are tailored to Heron's first stage mapping,
leaving a large space for performance improvement in Heron-like systems. 

In this paper, we focus on the two-staged instance placement problem that handles requests arriving in realtime. 
We target at highly modularized stream processing engines such as Apache Heron, to design an efficient instance placement scheme, with aligned objectives for both stages in minimizing traffic while maximizing resource utilization, so that the system can make timely but effective scheduling decisions. 
The challenges of the problem come from the combinatorial nature of mapping problems in both stages, the resource contention among different applications and their instances, the conflicts between objectives, 
and the online nature of request arrivals.

To address the challenges, various heuristics and approximation algorithms are available\cite{nemhauser1988integer}\cite{ma_framework}.
Notably, in recent years, there have been a progression of success in applying Monte Carlo Search Tree (MCTS) methods 
to solving problems that involve sequential decision processes \cite{browne2012survey}, such as the gameplay of Go\cite{silver2017mastering}. 
In fact, each mapping stage can also be viewed as a sequential decision process that places instances (containers) successively onto containers (servers). 
Moreover, MCTS takes the advantage of random sampling, achieving a tradeoff between computational efficiency and optimality largely affected by estimate accuracy.
To decide instance placements, low computational efficiency is favorable, 
while the accurate evaluation of consequent mappings is also critical. 
By leveraging MCTS, we propose MIPS, an efficient two-staged instance placement scheme for stream processing systems.
Our main results and contributions are as follows.

\textbf{Modeling and Formulation:}
With Apache Heron as the prototype, we develop a novel model that accurately  characterizes most updated stream processing systems with a highly modularized design. 
Based upon the model, we formulate the online instance placement problem as a two-staged constrained mapping problem, aiming at cross-server/container traffic reduction with high resource utilization. 
To our best knowledge, this is the first model for such stream processing systems and the first formulation for the two-staged mapping problem.

\textbf{Algorithm Design:} 
Considering the problem is in general $\mathcal{NP}$-hard, we adopt MCTS techniques with the Upper Confidence Bound for Trees (UCT), by taking the advantage of its \textit{anytime} property, \textit{i.e.}, more computing power generally leads to better performance.
By transforming each mapping stage to a sequential decision process, we propose MIPS, a randomized scheme that decides two-staged instance placement in a timely yet effective manner.
Besides, we refine MIPS from various aspects to promote its sampling quality, accelerate the computation, and overcome some practical issues.

\textbf{Experiment Verification and Analysis:}
Existing schemes are almost designed for Storm. To make them comparable with MIPS, we propose their variants under Heron. 
Then we conduct extensive simulations to evaluate MIPS against them.
Our results and analysis show that MIPS notably outperforms existing schemes with both low traffic and high utilization.

To the best of our knowledge, this is the first paper that studies the two-staged instance placement problem for Heron-like systems. 
The rest of the paper is organized as follows. Section II describes the model and problem formulation for the problem, while Section III proposes MIPS and discusses its variants to further accelerate the algorithm. 
Section IV presents results from simulations with the corresponding analysis, followed by the conclusion in Section V. 
\begin{center}
\begin{table}[!h]
\caption{Main Notations} \label{notations} 
\begin{tabular}{|c|l|}
\hline
\bf{Notation} & \bf{Description} \\
\hline
\hline
$\mathcal{S}$ & The set of servers in the cluster \\
\hline
$\Gamma_{s, k}$ & Capacity of $k$-th type of resource on server $s$ \\
\hline
\multirow{2}*{$\theta_{s, s'}$} & The communication cost per traffic unit from \\
& server $s$ to $s'$ \\
\hline
\multirow{2}*{$\mathcal{G}^{r}$} & Graph that corresponds to topology $r$ (speci- \\
& fied by request $r$) \\
\hline
$\mathcal{V}^{r}$ & The set of components in topology $r$ \\
\hline
$\mathcal{E}^{r}$ & The set of data streams in topology $r$ \\
\hline
$p(v)$ & The number of instances of component $v$ \\
\hline
$\mathcal{I}_{v}$ & The set of instances of component $v$ \\
\hline
$\mathbf{d}(i)$ & The resource demand of instance $i$ \\
\hline
$w(e)$ & The traffic rate of data stream $e$ \\
\hline
$\mathcal{C}^{r}$ & The set of containers of topology $r$ \\
\hline
$\mathbf{d}(c)$ & The resource capacity of container $c$ \\
\hline
\multirow{2}*{$\mathcal{C}(s, r)$} & The set of containers on server $s$ before \\
& deploying topology $r$ \\
\hline
{$\mathcal{I}^{r}$} & The set of all instances in topology $r$ \\
\hline
$X^{r}_{i, c}$ & Decision that places instance $i$ in container $c$ \\
\hline
$Y^{r}_{c, s}$ & Decision that places container $c$ on server $s$ \\
\hline
\end{tabular}
\end{table}
\end{center}

\section{Model and Problem Formulation}
We develop a model for stream processing systems based on Heron and formulate the two-staged instance placement problem. 
Main notations are summarized in Table \ref{notations}.

\subsection{Overall System Model}
We consider a Heron-based data stream processing system running within a cluster of servers, denoted by set $\mathcal{S}$. 
The servers are interconnected according to some network topology\cite{fat-tree}\cite{jellyfish}.
%, \textit{e.g.}, 
%Fat-Tree\cite{fat-tree} or Jellyfish\cite{jellyfish}. 
%The system evolves over time slots, indexed by $t$ $\in$ $\{0, 1, \dots\}$. Within time slot $t$, 
User requests arrive at the system in an online fashion, each denoted by $r$.
%$\mathcal{R}_{t}$.
Heron's scheduler receives and processes requests in a \textit{first-in-first-out} manner.
%Note that online processing is just a special case of this model by choosing a short enough slot length such that $|\mathcal{R}_{t}| \in \{0, 1\}$ for all $t$.
For each request $r$, 
%\in \mathcal{R}_{t}$
the scheduler instantiates its specified application and maps its instances to containers.
Next, the mapping is handed over to the underlying resource manager, \textit{e.g.}, YARN\cite{yarn} or Nomad\cite{nomad}. Then containers are assigned to servers, 
whereafter the application are set ready to run.
%Note that requests can be served in a batch or online manner. In a batch manner, the scheduler decides 

\subsection{Cluster Model}
Regarding each server, we consider $K$ types of resources, such as CPU cores, memory, and storage. The resource capacity of server $s$ is denoted by a vector $\boldsymbol{\Gamma}_{s}$, where $\Gamma_{s, k}$ denotes the capacity of $k$-th type of resource. 
Between any servers there always exists a non-blocking path for traffic transmission, which is commonly achievable in existing data center networks \cite{fat-tree}. 
Transmission from server $s$ to $s'$ incurs a communication cost of $\theta_{s, s'}$ per traffic unit, \textit{e.g.}, giga-bytes.

\subsection{Streaming Application Model}
Each request $r$ specifies the logical topology of a given application, denoted by a directed acyclic graph $\mathcal{G}^{r} \triangleq \left( \mathcal{V}^{r}, \mathcal{E}^{r} \right)$.
$\mathcal{V}^{r}$ denotes the set of components that make up the application, while $E^{r} \subseteq \mathcal{V}^{r} \times \mathcal{V}^{r}$ denotes the set of directed data streams between components. 
In practice, the diameter of $\mathcal{G}^{r}$ is often not too large, mostly less than four\cite{storm}. 

For component $v \in \mathcal{V}^{r}$, we use $p(v)$ to denote its parallelism, \textit{i.e.}, the number of its instances, and $\mathcal{I}_{v}$ to denote the set of its instances such that $|\mathcal{I}_{v}| = p(v)$.
The deployment of any instance $i \in \mathcal{I}_{v}$ has a resource demand of $\mathbf{d}(i) \in \mathbb{R}_{+}^{K}$.
For any instance $i$, we use $v(i)$ to denote its belonging component. 
  
For data stream $e \in \mathcal{E}^{r}$, its traffic rate is denoted by $w(e)$. In practice, traffic rates can be either pre-defined by users or estimated using historical data\cite{aniello2013adaptive}\cite{floratou2017dhalion}\cite{peng2015r}. 
Given any two components $v_1$ and $v_2$ such that $(v_1, v_2) \in \mathcal{E}^{r}$, the traffic is assumed evenly spread from instances of $v_1$'s to $v_2$'s, which is achievable by adopting shuffling policy in Heron or Storm\cite{storm}\cite{herondoc}. 
Therefore, the rate between instance $i_1 \in \mathcal{I}_{v_1}$ and $i_2 \in \mathcal{I}_{v_2}$ is obtained as $w(i_1, i_2) \triangleq \frac{w(v_1, v_2)}{p(v_1)p(v_2)}$.
For non-successive instances $i_1, i_2$, we set $w(i_1, i_2) = 0$.
The model can be easily extended to cases with uneven traffic patterns. 

\subsection{Deployment Model}
Besides logical topology, a request also specifies the number of containers to deploy instances. 
Per request $r$, the scheduler constructs a set of containers, denoted by $\mathcal{C}^{r}$. 
Each container $c \in \mathcal{C}^{r}$ has a resource capacity of $\mathbf{d}(c) \in \mathbb{R}_{+}^{K}$. 
Meanwhile, we denote the set of containers on server $s$ before deploying request $r$ by $\mathcal{C}(s, r)$. 

\subsection{Placement Decisions}
For request $r$, the instance placement consists of two stages. 

The first stage is to decide a mapping from instances of request $r$, \textit{i.e.}, $\mathcal{I}^{r} \triangleq \bigcup_{v \in \mathcal{V}^{r}} \mathcal{I}_{v}$ to containers $\mathcal{C}^{r}$, denoted by $\boldsymbol{X}^{r}$.
Each entry $X^{r}_{i, c} \in \{0, 1\}$ indicates whether instance $i$ is mapped to container $c$. The decision $\boldsymbol{X}^{r}$ maps each instance onto exactly one container, 
while ensuring the resource constraints for each container, \textit{i.e.},
\begin{equation}
	\begin{array}{l}\label{constraint: instance-container mapping}
		\displaystyle
		\sum_{c \in \mathcal{C}^{r}} X^{r}_{i, c} = 1, \ \forall i \in \mathcal{I}^{r} 
		\text{ and }
%	\end{array}
%\end{equation}
%\begin{equation}
%	\begin{array}{l}
		\displaystyle
		\sum_{i \in \mathcal{I}^{r}} 
		\mathbf{d}(i) \preceq \mathbf{d}(c),
		\ \forall c \in \mathcal{C}^{r}.
	\end{array}
\end{equation}

The second stage is to decide another mapping from containers $\mathcal{C}^{r}$ to servers $\mathcal{S}$, denoted by $\boldsymbol{Y}^{r}$.
Each entry $Y^{r}_{c, s} \in \{0, 1\}$ indicates whether container $c$ is mapped to server $s$.
The decision $\boldsymbol{Y}^{r}$ maps each container to exactly one server without violating the resource constraints on servers, 
\textit{i.e.},
\begin{equation}
	\begin{array}{l}\label{constraint: container-server-mapping}
		\displaystyle
		\sum_{c \in \mathcal{C}^{r}} 
		%% Across all containers or
		%% just for-each?
		Y^{r}_{c, s} = 1, \ \forall s \in \mathcal{S}
\\
		\displaystyle
		\sum_{c \in \mathcal{C}(s, r)}\!\!\!
		\mathbf{d}(c) +
		\sum_{c \in \mathcal{C}^{r}} Y^{r}_{c, s} \mathbf{d}(c) 
		\preceq \mathbf{\Gamma}_s, \ 
		\forall t \text{ and } s \in \mathcal{S}.
	\end{array}
\end{equation}

\subsection{Optimization Objectives}

Regarding instance-container mapping, 
it is highly desirable to place successive instances with data streams in between into the same containers 
in order to minimize cross-container traffic, reducing considerable communication overheads and shortening response time\cite{t-storm}.  
Formally, given decision $\boldsymbol{X}^{r}$ for request $r$, 
the total traffic between container $c, c' \in \mathcal{C}^{r}$ is 
\begin{equation}
	\begin{array}{c}
		\displaystyle
		T_{c, c'}(\boldsymbol{X}^{r}) \triangleq 
			\sum_{i, i' \in \mathcal{I}^{r}} 
			X^{r}_{i, c} X^{r}_{i', c'} 
			w(i, i').
	\end{array}
\end{equation}
Hence, the total cross-container traffic for request $r$ is 
\begin{equation}
	\begin{array}{l}
		\displaystyle
		T(\boldsymbol{X}^{r}) \triangleq 
		\sum_{c, c' \in \mathcal{C}^{r}} T_{c, c'}(\boldsymbol{X}^{r}).
	\end{array}
\end{equation}
On the other hand, deploying containers also incurs additional resource overheads\cite{docker}. 
For high resource utilization, the decision should map instances to as few containers as possible. 
Therefore, given decision $\boldsymbol{X}^{r}$, the number of utilized containers is 
\begin{equation}
	\begin{array}{l}
		\displaystyle
		U(\boldsymbol{X}^{r}) \triangleq
		\sum_{c \in \mathcal{C}^{r}} \min\{1, \sum_{i \in \mathcal{I}^{r}} X^{r}_{i, c}\},
	\end{array}
\end{equation}
where the term for container $c$ is equal to one only if there is any instance residing in container $c$.

Regarding container-server mapping, containers with intensive traffic in between should be placed closely to minimize the inter-server communication cost. 
Fixed $\boldsymbol{X}^{r}$ and given $\boldsymbol{Y}^{r}$, the cost for request $r$ between server $s$ and $s'$ is 
\begin{equation}
	\begin{array}{l}
		\displaystyle
		W_{s, s'}(\boldsymbol{Y}^{r}) \triangleq
		\sum_{c, c' \in \mathcal{C}^{r}} 
			Y^{r}_{c, s} Y^{r}_{c', s'} \theta_{s, s'} 
			T_{c, c'}(\boldsymbol{X}^{r}).
	\end{array}
\end{equation}
The total communication cost incurred after deployment is
\begin{equation}
	\begin{array}{l}
		\displaystyle
		W(\boldsymbol{Y}^{r}) \triangleq
			\sum_{s, s' \in \mathcal{S}} W_{s, s'}(\boldsymbol{Y}^{r}).
	\end{array}
\end{equation}

\subsection{Problem Formulation}
For request $r$, we formulate the instance-container mapping problem ({\bf ICMP}) as
\begin{equation}
	\begin{array}{cl}\label{Prob-def: icmp}
		\underset{\boldsymbol{X}^{r}}{\text{Minimize}}
		& \displaystyle
		\alpha T(\boldsymbol{X}^{r}) + (1-\alpha) U(\boldsymbol{X}^{r}) \\
		\text{Subject to} & 
		\displaystyle
		X^{r}_{i, c} \in \{0, 1\}
		\text{ and } (\ref{constraint: instance-container mapping}).
	\end{array}
\end{equation}
where $\alpha \in [0, 1]$ is a tunable parameter that weights the importance of cross-container traffic reduction compared to decreasing the number of utilized containers. 
Meanwhile, we define the container-server mapping problem ({\bf CSMP}) as
\begin{equation}
	\begin{array}{cl}\label{Prob-def: csmp}
		\underset{\boldsymbol{Y}^{r}}{\text{Minimize}}
		& \displaystyle
		W(\boldsymbol{Y}^{r}) \\
		\text{Subject to} 
		& \displaystyle
		Y^{r}_{c, s} \in \{0, 1\}
		\text{ and } (\ref{constraint: container-server-mapping}).
	\end{array}
\end{equation} 

\section{Algorithm Design}
{\bf ICMP} and {\bf CSMP} are both non-linear combinatorial optimization problems. 
Such problems are generally $\mathcal{NP}$-hard with a huge search space size (\textit{e.g.}, $O(2^{|\mathcal{I}^{r}| \times |\mathcal{C}^{r}|})$ for {\bf ICMP}), while coupled resource constraints add even more complexity.
Inspired by recent progressive success\cite{silver2017mastering} in applying random sampling methods like MCTS to solving complex problems with sequential decision making processes,  
we shift our perspective by viewing each stage as a sequential decision  process that places instances (or containers) successively. 
By leveraging MCTS, we aim at developing an efficient scheme to solve each stage of mapping, hopefully well balancing the computational complexity and effectiveness.

\subsection{Overview of MCTS}
MCTS is derived from tree search methods\cite{russell2016artificial} that handle sequential decision processes.
The key idea of such tree search methods is to build a single-rooted tree that corresponds to the process. 
Each tree node represents a system state, while each outgoing edge  represents the action that leads to the next state. 
In this way, each path from root node to a leaf indicates a complete decision sequence with an eventual reward. 
The problem then turns to be finding a policy that chooses an action to execute from current node to the next node with maximum reward.  
In some cases, direct construction of the search tree may require excessive compute resource due to its large search space size and branching factor.

MCTS takes a detour by incrementally constructing part of the tree with random sampling on a round basis.
Each node maintains the estimates of the reward of executing different actions to its child nodes.
Within each round, MCTS proceeds by following a framework with four basic steps: 
1) \textit{Traversal}: Starting from the root node, recursively finds the next node to traverse by choosing the one with the maximum reward estimate, until reaching a leaf node or some unexpanded node (with unvisited child nodes);
2) \textit{Expansion}: Upon an unexpanded node, MCTS expands it by adding one of its unvisited child to the partially built tree;
3) \textit{Simulation}: Starting from a newly added node, MCTS conducts a random simulation to sample a complete decision sequence.
4) \textit{Back-propagation}: The reward induced by the acquired sequence is then back-propagated along the way to the root node, refining the reward estimates of visited nodes.
After numbers of rounds (samples), MCTS returns the next action from the root to the next child node that is most likely towards the optimal decision sequence.
The process is then replayed on the subtree that is rooted at the chosen child node. 
\begin{algorithm}
 \caption{MIPS for the two-staged instance placement}
 \begin{algorithmic}[1]
 \label{algo}
 \STATE \%\% Handle request $r$.
 \STATE \textbf{function} MIPS\_FOR\_ICMP($\mathcal{I}^{r}$, $\mathcal{C}^{r}$)
 \STATE $~~$ Initialize $\text{count} \leftarrow 0$ and action sequence $\mathcal{A} \leftarrow \emptyset$.
 \STATE $~~$ Initialize root node $\eta_{root}$ with $\mathcal{C}^{r}$.
 \STATE $~~$ \textbf{while} $\text{count} < |\mathcal{I}^{r}|$ \textbf{do}
 \STATE $~~~~$ $(a, \eta) \leftarrow \text{NEXT\_ACTION}(\eta_{root}, \text{stage}\!=\!1)$
 \STATE $~~~~$ Update $\mathcal{A} \leftarrow \mathcal{A} \bigcup \{ a \}$ and $\eta_{root} \leftarrow \eta$
 \STATE $~~~~$ $\text{count} \leftarrow \text{count} + 1$
 \STATE $~~$ \textbf{return} $\mathcal{A}$
 \STATE
 \STATE \textbf{function} MIPS\_FOR\_CSMP($\mathcal{C}^{r}$, $\mathcal{S}$)
 \STATE $~~$ Initialize $\text{count} \leftarrow 0$ and action sequence $\mathcal{A} \leftarrow \emptyset$.
 \STATE $~~$ Initialize root node $\eta_{root}$ with $\mathcal{S}$.
 \STATE $~~$ \textbf{while} $\text{count} < |\mathcal{C}^{r}|$ \textbf{do}
 \STATE $~~~~$ $(a, \eta) \leftarrow \text{NEXT\_ACTION}(\eta_{root}, \text{stage}=2)$
 \STATE $~~~~$ Update $\mathcal{A} \leftarrow \mathcal{A} \bigcup \{ a \}$ and $\eta_{root} \leftarrow \eta$
 \STATE $~~~~$ $\text{count} \leftarrow \text{count} + 1$
 \STATE $~~$ \textbf{return} $\mathcal{A}$
 \end{algorithmic}
\end{algorithm}
\setlength{\textfloatsep}{0pt}

\begin{algorithm}
 \caption{Sub-functions for MIPS}
 \begin{algorithmic}[1]
 \label{algo2}
% \STATE \%\% Decide the next mapping action to take from node $\eta$.
 \STATE \textbf{function} NEXT\_ACTION($\eta$, sid) \\
 \STATE $~~$ Set $\eta$ as root node $\eta_{0}$.
 \STATE $~~$ Initialize $t \leftarrow 0$ and MAX\_SAMPLE\_NUM.  
 \STATE $~~$ \textbf{while} $t <$ MAX\_SAMPLE\_NUM \textbf{do}
 \STATE $~~~~$ $\eta \leftarrow \text{TRAVERSE}(\eta_{0})$
 \STATE $~~~~$ $\Delta \leftarrow \text{SIMULATE}(\eta, \text{stage}=\text{sid})$
 \STATE $~~~~$ \textbf{if} $\Delta > 0$ \textbf{then}
 \STATE $~~~~~~$ $\text{BACK\_PROP}(\eta, \Delta)$
 \STATE $~~~~~~$ $t \leftarrow t + 1$ and 
% \STATE $~~~~~~$ 
 $\text{best\_child} \leftarrow \text{BEST\_CHILD}(\eta_{0}, 0)$
 \STATE $~~$ \textbf{return} 
 	$a(\eta_{0}, \text{best\_child})$, $\text{best\_child}$
 \STATE
 \STATE \textbf{function} TRAVERSE($\eta$)
 \STATE $~~$ \textbf{while} $\eta$ is not a leaf \textbf{do}
 \STATE $~~~~$ \textbf{if} $\mathcal{A}_{\text{untried}}(\eta) \neq \emptyset$ \textbf{then}
 \STATE $~~~~~~$ \textbf{return} EXPAND($\eta$) 
 \STATE $~~~~$ \textbf{else}
% \STATE $~~~~~~$ 
 $\eta \leftarrow \text{BEST\_CHILD}(\eta, \omega=\sqrt{2})$
 \STATE $~~$ \textbf{return} $\eta$
 \STATE
 \STATE \textbf{function} EXPAND($\eta$)
 \STATE $~~$ Choose $(i_0, c_0) \in \mathcal{A}_{\text{untried}}(\eta)$ uniformly randomly
 \STATE $~~$ Place instance $i_0$ onto container $c_0$
 \STATE $~~$ \textbf{return} the resultant node $\eta'$
 \STATE
 \STATE \textbf{function} BEST\_CHILD($\eta$, $\omega$)
 \STATE $~~$ \textbf{return} $\underset{\eta' \in M(\eta)}{\arg\min} 
 	\frac{Q(\eta')}{N(\eta')+1} - \omega \sqrt{\frac{
 		2 \ln N(\eta)}{N(\eta')}}
 $ 
 \STATE
 \STATE \textbf{function} SIMULATE($\eta$, sid)
 \STATE $~~$ \textbf{while} $\eta$ is not a leaf \textbf{do}
 \STATE $~~~~$ 
 	Choose $a \in \mathcal{A}_{\text{untried}}(\eta)$ uniformly randomly 
 \STATE $~~~~$ Execute $a$ and obtain the resultant node $\eta'$
 \STATE $~~~~$ Set $\eta \leftarrow \eta'$
 \STATE $~~$ \textbf{if} in stage $sid$ and $\eta$ satisfies constraints ($sid$)
%	 \ref{constraint: container-server-mapping})
 	\textbf{then}
 \STATE $~~~~$ \textbf{return} $\Delta(\eta)$  \%\% Reward at leaf node $\eta$.
 \STATE $~~$ \textbf{else}
% \STATE $~~~~$ 
 \textbf{return} $-1$ \%\% Ends in an invalid mapping.
 \STATE
 \STATE \textbf{function} BACK\_PROP($\eta$, $\Delta$)
 \STATE $~~$ \textbf{while} $\eta$ is not root \textbf{do}
 \STATE $~~~~$ $N(\eta) \leftarrow N(\eta) + 1$ and
% \STATE $~~~~$ 
 $Q(\eta) \leftarrow Q(\eta) + \Delta$
 \STATE $~~~~$ 
 $\eta \leftarrow$ parent of $\eta$
\end{algorithmic}
\end{algorithm}
\setlength{\textfloatsep}{0pt}

MCTS has various favorable properties that contribute to its successful application. One of them is its \textit{anytime} property. 
For an anytime algorithm, it can return a valid solution whenever interrupted before it ends; on the other hand, more compute resources generally leads to results of better quality, well balancing the tradeoff between computational complexity and optimality of solution.
This property is particularly desirable for instance placement: 
On one hand, given its overwhelming search space size and progressively request arrivals, 
the scheduler should determine an effective placement but in a timely fashion, instead of undertaking a time-consuming decision process; 
on the other hand, provided with more resources, the scheduler should be able to improve the placement rather than resort to tedious parameter tuning tricks. 
Another is that MCTS only provides a generic framework, whereby system designers can customize the basic steps to further optimize their applications.

\subsection{Modeling Decision Trees and Algorithm Design}
To leverage MCTS, we need to cast each mapping stage into a sequential decision process.

\textbf{Decision Tree for ICMP}: 
First, we transform the decision process for instance-container mapping into a decision tree.
For request $r$, we construct a tree for its instance-container mapping, with each node $\eta$ denoting the state associated with a given mapping. 
The root node $\eta_{root}$ corresponds to the state where no instances are mapped to containers, \textit{i.e.},
$X^{r}_{i, c} = 0$ for all $i$ and $c$.
Each outgoing edge of $\eta_{root}$ indicates an action of mapping some unmapped instance to some container, subject to the resource constraint in (\ref{constraint: instance-container mapping}).
For example, the action of mapping instance $i_0$ to container $c_0$ changes the mapping state from $\eta_{root}$ to $\eta$ with only $X^{r}_{i_0, c_0} = 1$, denoted by $a(\eta_{root}, \eta)$.
Similarly, its child nodes then point to their descendants. 
Recursively defined in this way, the tree eventually reaches leaf nodes with mapping that satisfy (\ref{constraint: instance-container mapping}).
The reward $\Delta$ for each leaf node is set as the associated objective value defined in (\ref{Prob-def: icmp}) given its mapping.

\textbf{Decision Tree for CSMP}:
Regarding CSMP, for request $r$, we construct another tree for container-server mapping. 
The root corresponds to the state where no given containers of request $r$ are mapped to servers, while each of its outgoing edge denotes the action of mapping some unmapped container to some server subject to resource constraint in (\ref{constraint: container-server-mapping}). 
Edges then point to its child nodes with resultant mapping, which in turn point to more descendants. 
Leaf nodes are either valid mappings from containers to servers, or mappings  interrupted due to limited resource.
The reward for each leaf node is the corresponding objective value defined in (\ref{Prob-def: csmp}) given its mapping.

Every node $\eta$ in the above trees maintains four states. 
The first is $N(\eta)$, denoting the times of node $\eta$ being visited.
The second is $Q(\eta)$, denoting the total accumulated reward node $\eta$ has received so far.
In this way, $\frac{Q(\eta)}{N(\eta)}$ reflects the expected reward induced by following the decision sequence through node $\eta$.
The third is $\mathcal{A}_{\text{untried}}(\eta)$, denoting the set of all untried mapping actions that satisfy resource constraint in (\ref{constraint: instance-container mapping}). 
The last is $M({\eta})$, denoting the set of $\eta$'s children being visited. 

To find the optimal decision sequence for each decision tree,
we propose \textit{MIPS}, \textit{i.e.}, \textit{M}TCS-based \textit{IN}stance placemen\textit{T} scheme to decide two stages of mapping, respectively.
Algorithm \ref{algo} shows the pseudocode of MIPS that decides the two-staged instance placement for each request $r$. 

Notably, to choose the best child node (Alg.\ref{algo2}, line 25),  
MIPS has to decide: 1) to exploit historical information by choosing from the visited ones with the minimum objective value, or 2) to explore an unvisited node with unknown reward, \textit{a.k.a.} the \textit{exploitation-and-exploration} tradeoff\cite{browne2012survey}. 
To find the best possible placement, MIPS must well balance the tradeoff since 
1) if over-dependent on the historical information, MIPS may miss unvisited nodes that lead to better placement, while 2) radical exploration might waste resources on nodes with a far worse reward.  
To this end, MIPS leverages the widely adopted \underline{U}pper \underline{C}onfidence bound for \underline{T}rees (UCT)\cite{kocsis2006improved}. 
By viewing the problem as a multi-armed bandit problem\cite{kocsis2006bandit},
UCT chooses the best child node with the minimum upper confidence bound 1 (UCB1) value (line 25 of Alg.\ref{algo2})\cite{browne2012survey}.  
The left-hand-side term of UCB is the reward estimate, obtained by averaging the aggregating rewards from past samples (to ensure the term to be finite, we add one to the denominator); the right-hand-side reflects the visited frequency. 
If a node has never been visited, the term goes to infinity and its UCB value is $-\infty$, thus the node must be chosen with precedence. If a node is rarely visited but its parent node is visited a great number of times, the node will have a higher chance to be chosen.
In this way, MIPS can rebalance the tradeoff by choosing a proper value of weight parameter $w$; thus greater $w$ induces a more explorative search.

\subsection{System Workflow}
Upon request $r$'s arrival, the system first parses its topology, resource demand, and parallelism requirement. 
In the first stage, the system scheduler applies MIPS to obtain the instance-container mapping.
Then it eliminates the containers with no instances assigned and submits each container as a job to the underlying resource manager \cite{nomad}. 
In the second stage, cluster scheduler applies MIPS to decide the mapping from containers to servers and enforces the deployment. 
In practice, MIPS can be implemented as a custom module through APIs provided by Heron and cluster schedulers \cite{herondoc}\cite{yarn}\cite{nomad}.

\subsection{Practical Issues and Refinement}
By leveraging random sampling, 
the effectiveness of MIPS heavily depends on the estimate accuracy for the objective values of the tree nodes (states). 
Accurate estimates often require a large number of samples, inducing long computational time and massive compute resources.
Considering that some samples may lead to decision sequences with unfavorably high objective values, 
uniformly random simulation and node expansion may still have much room for improvement. 
Hence, to promote sampling quality, we optimize MIPS by
refining its way of 1) selecting unvisited children, 2) choosing the best child, and 3) simulating.  

\textbf{1) Expansion policy 
        (Alg.\ref{algo2}, line 19-22):} 
        For {\bf ICMP}, given node $\eta$ with unvisited children (untried actions), instead of uniformly random selection, we favor the action that places an instance to such a container that hosts any of its successive instances. 
We assign such actions with a positive score and the rest with zero score, while selecting an action only from those with high scores.
Thus MIPS biases the mappings that place successive instances in proximity, reducing cross-container traffic with least resources. 
Regarding {\bf CSMP}, MIPS favors mapping containers with traffic in between on the same server. 
 
\textbf{2) Best child selection
	(Alg.\ref{algo2}, line 24-25):}
	Although UCT well balances the exploration-and-exploitation tradeoff, 
	MIPS must explore all children nodes before re-passing the visited ones. 
	However, some actions may obviously lead to high objective values, as discussed previously. 
	We bias such node $\eta$ by initializing $N(\eta) = 1$ and $Q(\eta)$ with a large positive value, pretending that the node has been visited once \textit{a priori} with an unfavorably high objective value. 

\textbf{3) Simulating (Alg.\ref{algo2}, line 27-34):} Starting from a given node, MIPS simulates the rest of mapping decision process by repetitively  choosing an unmapped instance uniformly randomly and assigning it to one of the containers. 
However, such an aimless policy may lead to decision sequences that place successive instances to different containers,
incurring undesirably considerable traffic and resource overheads. 
Instead, we refine the simulating strategy in the following way: Each time  MIPS randomly chooses one of the unmapped instances and maps it to the container with minimum incremental cross-container traffic. 
Likewise, when applied to \textbf{CSMP}, MIPS simulates by progressively placing the rest unmapped containers to servers with minimum incremental cross-server traffic. 
%With the above refinements, we can promote the sampling quality and obtain better mapping decisions with less samples. 

\begin{figure*}[t!]
	\centering
	\includegraphics[scale=.2]{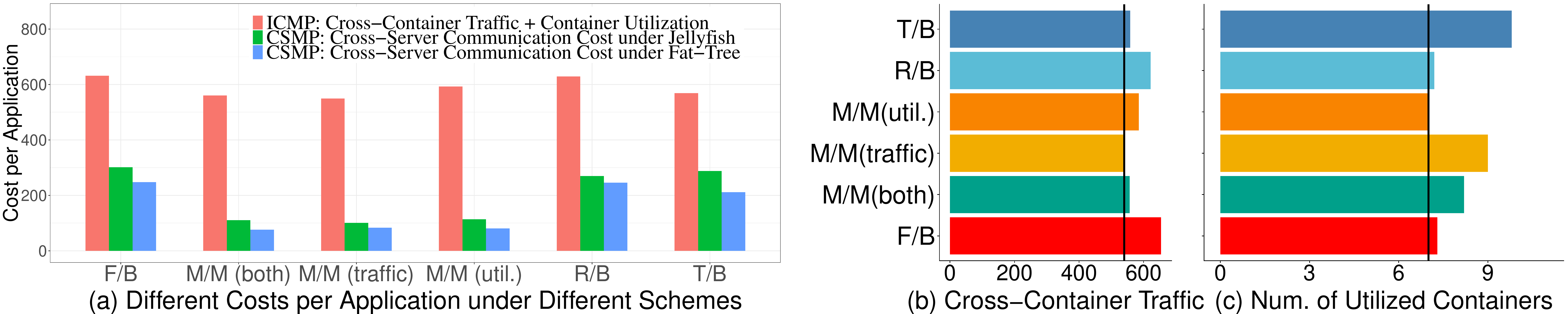}
	\caption{
	   Comparisons between MIPS/MIPS (M/M) and other schemes: R-Heron/Best-fit (means R-Heron in the first stage and Best-fit in the second stage), T-Heron/Best-fit, and FFD/Best-fit, denoted by R/B, T/B, F/B, respectively. 
	   We run M/M under different values of $\alpha$, including $\alpha=1$ (\textit{M/M traffic}: minimizing traffic only), $\alpha=0$ (\textit{M/M util.}: minimizing container utilization only), and $\alpha=0.5$ (\textit{M/M both}: targeting both objectives), according to (\ref{Prob-def: icmp}).
	}
	\label{fig-comparisons}
\end{figure*}
\setlength{\textfloatsep}{0pt}

\section{Simulation}

\subsection{Basic Settings}

\textbf{Cluster Topology:}
We prototype a stream processing system based on Heron\cite{herondoc} and cluster resource manager Nomad\cite{nomad}.
We implement two custom schedulers based on MIPS in the system and Nomad for instance-container and container-server mapping, respectively.
The system is deployed in clusters that are constructed using two widely adopted topologies, 
Jellyfish\cite{jellyfish} and Fat-Tree\cite{fat-tree}, respectively.
Within each cluster are $24$ homogeneous switches and $16$ heterogeneous servers. 
Each switch has a port number of $4$, with a bandwidth of $40$Gbps on each port.
For any two servers, the unit communication cost of transferring data streams is set as the number of hops of the shortest path between them.

\textbf{Deployment Resources:}
Regarding resource allocation, we consider CPU cores and memory on 
servers\cite{herondoc}. 
Every server has a number of CPU cores ranging from $16$ to $64$ and memory from $8$G to $32$G.
For each stream processing application, all of its containers have identical resource capacities. 
  
\textbf{Stream Processing Applications:}
We progressively submit requests to the system scheduler to deploy applications with common topologies\cite{kulkarni2015twitter}\cite{t-storm}\cite{peng2015r}.
Each request specifies a topology with a depth varying from $3$ to $5$, and a number of components ranging from $3$ to $6$.
Besides, the parallelism for each component ranges from $2$ to $6$.
Instances of the same component have identical functionalities. 
Instances' resource demand varies from $2$ to $6$ CPU cores and $4$ to $8$GB memory. 

%\textbf{Traffic Workloads:}
%We conduct trace-driven simulations, where tuple arrivals follow the measurements that are drawn from real-world network systems\cite{benson2010network}.  
%In addition, we also conduct simulations where tuple arrivals follow Poisson distribution, with the same arrival rate as in trace-driven cases. 

\textbf{Compared Schemes:}
Besides Heron's first-fit-decreasing (FFD) scheme, most existing schemes are designed for Storm \cite{peng2015r}\cite{t-storm}.
To make them comparable with MIPS, we propose their variants for instance-container mapping under Heron.

\textit{R-Heron}: 
Given an application, initialize all its containers. 
Enumerate its components by a breadth first traversal on its topology. 
If the topology has more than one sink node, then 
add a virtual root node that precedes all sink nodes and apply the traversal.
Next, for each component, enumerate its instances and repeat the following process. 
For each instance, assign it to the container with minimum resource distance,
where the distance is defined as the traffic rate between the container and other containers, adding the euclidean distance between the instance's resource demand vector and the container's available resource vector.
If no containers have enough resources to host an instance, then an error will be raised.

\textit{T-Heron}:
Given an application, initialize all its containers.
Sort all instances by their descending order of (incoming and outgoing) traffic rate. 
Then assign each instance to one of its application's containers 
with minimum incremental traffic and without exceeding the resource capacity of the container. 

\textit{FFD}\cite{herondoc}:
Given an application, initialize all its containers, an empty active container list, and a list of unmapped instances. 
While there still exists unmapped instances, 
repeat the process: 1) Choose the next unmapped instance from the list; 
2) sort the active containers by descending order of their available resources;
3) pick the first active container with sufficient resource; if no active container can host the instance, add a new container to the list and assign the instance to it. 

We adopt the best-fit scheme in Nomad\cite{nomad} as the underlying container-server mapping scheme for the baselines, which assigns each container to the server with free resources that best match its resource demand. 

\subsection{Results and Analysis}
We show and analyze the results from our extensive simulations of MIPS. Since MIPS is a randomized algorithm, we repeat each simulation for $100$ times and take the average of the results to eliminate the impact of randomness. 

\textbf{Performance against Other Schemes:} 
Figure \ref{fig-comparisons} compares two-staged MIPS (M/M) with other three schemes in terms of costs in two stages. The number of samples per round is fixed as $500$ for MIPS.  
Figure \ref{fig-comparisons} (a) makes a comparison of total costs in each of the two stages induced by different schemes, respectively. 
Note that for any scheme, the total cost of ICMP remains the same while only the cost of CSMP differs under Fat-Tree and Jellyfish, 
since the decision making for ICMP does not involve communication costs that vary in topologies.
We make the following observations. 

\textit{In the first stage}, MIPS (M/M) with different values of $\alpha$ effectively reduces the total cost of cross-container traffic and container utilization compared to other schemes. For example, M/M (traffic) with $\alpha\!\!=\!\!1$ leads to the minimum cost of $549.152$, with a $13\%$ reduction to F/B, $4\%$ to T/B, and $12\%$ to R/B. 
M/M (util) and M/M (both) also lead to cost reduction but slightly less than inferior to M/M (traffic). 

Zooming into the observation, we further compare the cross-container traffic and container utilization in Figure \ref{fig-comparisons} (b) and (c), respectively. 
Figure \ref{fig-comparisons} (b) shows that MIPS (traffic) incurs the minimum cross-container traffic. This is reasonable since with $\alpha=1$, MIPS assigns successive instances into the same containers in the best way possible. 
Different from THeron also with a relatively low traffic, MIPS decides the placement based on its experiences acquired from random sampling and evaluation rather than greedy heuristics, leading to less traffic. 
Meanwhile, other heuristics RHeron and FFD bring more traffic due to their less or no focus on traffic reduction.

Figure \ref{fig-comparisons} (c) shows that M/M (util.) achieves the minimum container utilization at $7$. Meanwhile, M/M (util.) also outperforms heuristics FFD and RHeron that focus on utilization by $6\%$ less traffic.
On the other hand, along with the traffic reduction, extra container utilization comes as a price to M/M (traffic) for its traffic reduction. 
Though, M/M (traffic) still outperforms THeron in container utilization.  
Jointly considering traffic reduction and utilization, M/M (both) make a well balance with about $2\%$ traffic increase and little extra utilization to the optimum at both sides.
All such advantages are conduced by MIPS's well exploitation of random sampling. 

\textit{In the second stage}, Figure \ref{fig-comparisons} (a) shows that M/M significantly outpaces other schemes by an up to $60\%$ reduction in cross-server traffic under both topologies. This ascribes to not only the advantage taken from the mapping in the first stage, but the effectiveness of MIPS in the second stage as well.

\begin{figure}[!t]
	\centering
	\subfigure{
		\includegraphics[scale=.21]{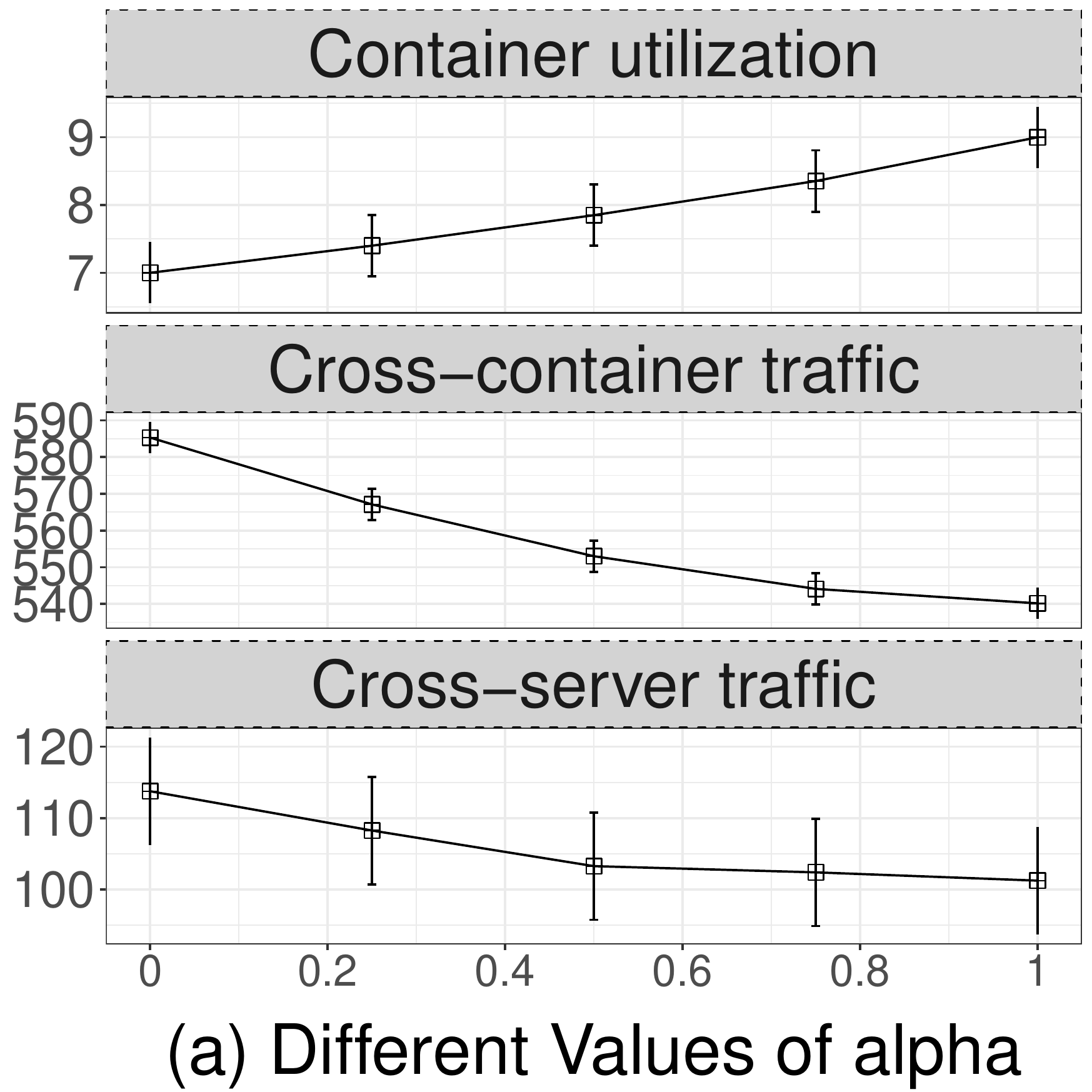}
	}
	\subfigure{
		\includegraphics[scale=.21]{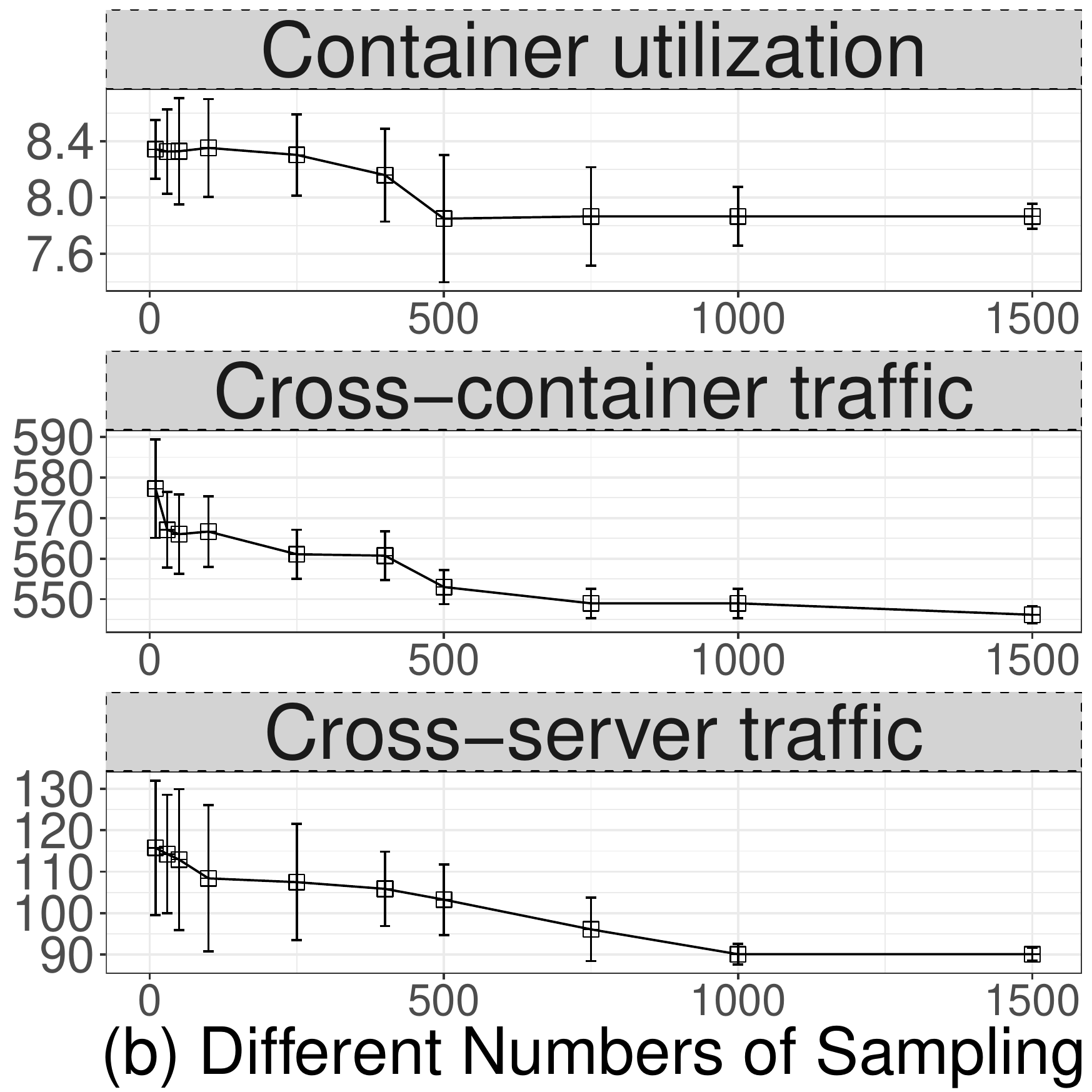}
	}
	\caption{
	   MIPS's Performance sunder various choices of $\alpha$ and sample numbers. 
	}
	\label{alpha_iter}
\end{figure}

\textbf{Performance under Different Values of $\alpha$:}
From previous results, there seems to be a potential tradeoff between cross-container traffic and container utilization. Figure \ref{alpha_iter} verifies the relationship qualitatively by showing the costs incurred by MIPS under Fat-Tree with $\alpha$ growing from $0$ to $1$: container utilization gently increases while cross-container traffic notably lessens, with cross-server traffic  decreasing as well. 
It seems plausible to place successive instances close to each other to reduce cross-container traffic with only a few containers.
However, due to the heterogeneity of instance resource demands, this intuition may fail, as exemplified previously by Figure \ref{example1} (a). 
Moreover, with online requests arrivals, prior deployed applications may take up major server resources, leaving only fragmented resources on servers for later arriving applications. 
Their containers would either be deployed on new servers or placed distantly on across existing servers. 
MIPS carefully places instances to minimize the impact of dependence between successive placement decisions. 

\setlength{\textfloatsep}{0pt}
\begin{figure}[!t]
	\centering
	\subfigure{
		\includegraphics[scale=.21]{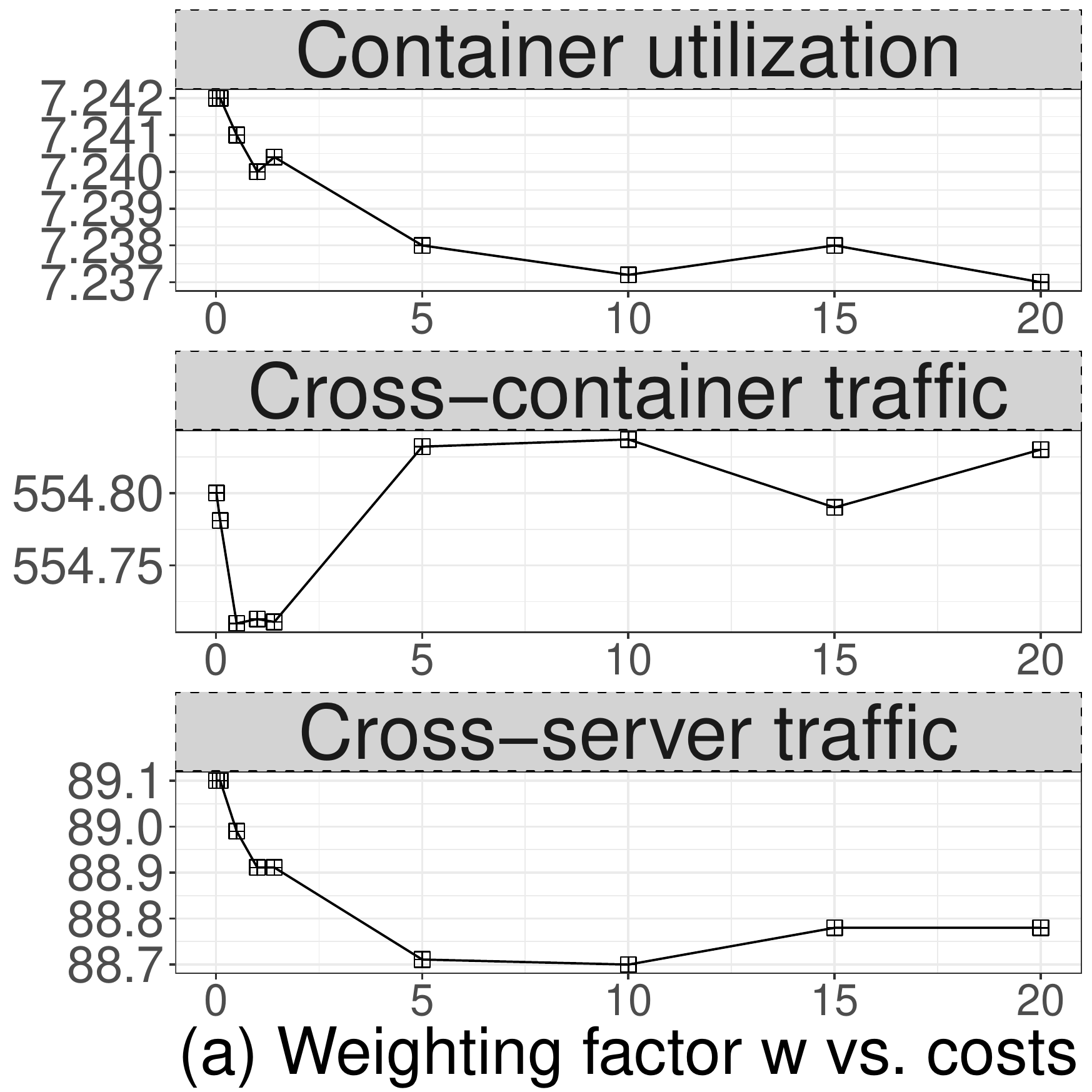}
	}
	\subfigure{
		\includegraphics[scale=.21]{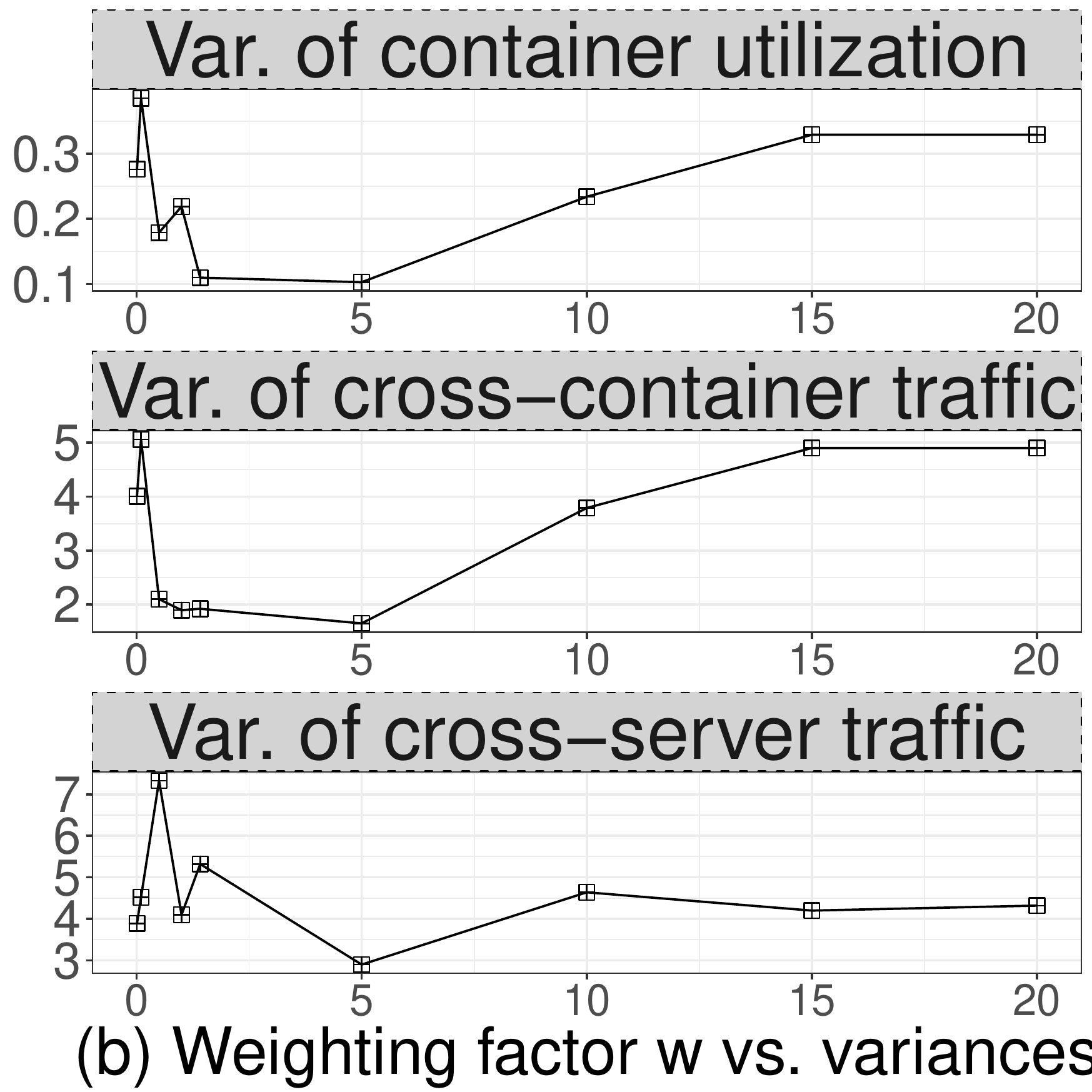}
	}
	\caption{
	   MIPS's Performance sunder various choices of $\alpha$ and sample numbers. 
	}
	\label{weighting factor}
\end{figure}
\setlength{\textfloatsep}{0pt}

\textbf{Performance with Different Sampling Numbers:} Besides, sampling number is another key factor to MIPS's performance --
more sampling means simulating more possible sequences, conducing to more accurate estimates for MIPS's eventual decision making. However, that also requires longer computational time and more compute resources being consumed. 
A natural question is that how many samples are sufficient to decide placements with low traffic and resource consumption. 
Figure \ref{alpha_iter} investigates the relationship between sampling number and MIPS's performance, with $\alpha=0.5$. 
As the sampling number grows from $0$ to $500$, there is a significant reduction in the container utilization and cross-container traffic. However, as the sampling number continues to rise, the improvement gradually fades and eventually converges at around $1000$ samples. 
This implies that in practice, compared to its enormous sample space size, MIPS requires only mild-value of sampling number to make timely yet efficient decisions with effective placement with both low traffic and few container resources. 

\textbf{Performance under Different Exploration-Exploitation Tradeoffs:} 
To find the best possible placement within a limited number of samples, MIPS has to decide in each round either to exploit decision sequences with known reward estimates or explore those with unknown rewards. 
Figure \ref{weighting factor} investigates the impact of weighting parameter $w$ on the system performance and the variance among repeated simulations, with $\alpha=0.5$ and sample number as $500$ per round under Fat-Tree topology.  
Figure \ref{weighting factor} (a) shows that as parameter $w$ varies from $0$ to $20$,
MIPS incurs costs roughly at the same level, with container utilization around $7.239$, cross-container traffic around $554.79$, and cross-server traffic around $88.8$.
This is reasonable since with fixed settings those costs are supposed to remain constant on the long-term average.
However, we can still see a slight fluctuation among the results under different choices of $w$. The reason lies in the sampling quality induced by different exploration-exploitation tradeoffs being made. 
With a smaller value of $w$, MIPS tends to exploit those decision sequences with known estimates, 
making the resultant decision largely dependent on a limited set of sequences while missing those with unknown but potentially better rewards. 
On the other hand, a greater value of $w$ leads to a more explorative search.
%: 
%MIPS tends to explore those sequences that have rarely or never been visited. This might result in samples with greater or far worse rewards. 
Due to the randomness of sampling, either undue exploitative or explorative search can have a large variance among different simulations. 
Figure \ref{weighting factor} (b) verifies this: The variance of the three costs all rise up after a drop from $w=0$ to $5$, suggesting that $w=5$ is a proper value for MIPS to balance the trade-off.

\section{Conclusion}
In this paper, we studied the two-staged instance placement problem for stream processing engines like Heron. 
By modeling each stage as a sequential decision-making process and leveraging MCTS to the problem,
we proposed MIPS, a randomized scheme that decides the instance placement in two stages in a timely yet efficient manner. 
To promote the sampling quality, we refined MCTS from various aspects and discussed practical issues. 
To evaluate MIPS against existing schemes, we propose variants of the schemes in Heron-like systems. 
Results from extensive simulations show that MIPS outperforms existing schemes with both low traffic and high resource utilization, 
but requires only mild-value of sampling.

\end{document}